\documentclass[fleqn,usenatbib]{mnras}

\usepackage[T1]{fontenc}
\usepackage{graphicx}
\usepackage{amsmath}
\usepackage{url}
\hypersetup{hypertexnames=false}

\DeclareRobustCommand{\VAN}[3]{#2}
\let\VANthebibliography\thebibliography
\def\thebibliography{\DeclareRobustCommand{\VAN}[3]{##3}\VANthebibliography}

\title[Radio evolution of Swift~J1818.0$-$1607]{Extended 2.25/8.60~GHz Observations of the Magnetar Swift~J1818.0$-$1607}

\author[Huang et al.]{
Zhi-Peng Huang,$^{1,2}$
Zhen Yan,$^{3,4,5}$\thanks{E-mail: yanzhen@shao.ac.cn}
Zhi-Qiang Shen,$^{3,4,5}$\thanks{E-mail: zshen@shao.ac.cn}
Hao Tong,$^{6}$
Wei-Yang Wang,$^{7}$\newauthor
Xiao-Yu Lai,$^{1,2}$
Rong-Bing Zhao,$^{3,5}$
Ya-Jun Wu,$^{3,5}$
Rui Wang,$^{8}$
Xiao-Wei Wang,$^{3,4}$\newauthor
Fan Yang,$^{3,9}$
Chu-Yuan Zhang,$^{3,4}$
Zhen-Long Liao,$^{3,4}$
and Yang-Yang Lin$^{3,10}$
\\
$^{1}$School of Optoelectronic Information Engineering, Hubei University of Education, Wuhan 430205, China\\
$^{2}$Institute of Astronomy and High Energy Physics, Hubei University of Education, Wuhan 430205, China\\
$^{3}$Shanghai Astronomical Observatory, Chinese Academy of Sciences, Shanghai 200030, China\\
$^{4}$School of Astronomy and Space Sciences, University of Chinese Academy of Sciences, Beijing 100049, China\\
$^{5}$State Key Laboratory of Radio Astronomy and Technology, Shanghai Astronomical Observatory, Chinese Academy of Sciences, Shanghai 200030, China\\
$^{6}$Department of Astronomy, School of Physics and Materials Science, Guangzhou University, Guangzhou 510006, China\\
$^{7}$University of Chinese Academy of Sciences, Beijing 100049, China\\
$^{8}$School of Information Engineering, Sanming University, Sanming 365004, China\\
$^{9}$School of Physical Science and Technology, ShanghaiTech University, Shanghai 201210, China\\
$^{10}$College of Science, Shanghai University, Shanghai 200444, China
}

\date{Accepted 2026 November 2. Received 2026 November 1; in original form 2026 July 28}
\pubyear{2026}

\begin{document}
\label{firstpage}
\pagerange{\pageref{firstpage}--\pageref{lastpage}}
\maketitle

\begin{abstract}
We present the long-term rotational evolution and radio emission properties of the magnetar Swift~J1818.0$-$1607, using simultaneous 2.25/8.60~GHz observations obtained with the Shanghai Tian Ma Radio Telescope from MJD~58936 to 59427. The average pulse profiles exhibit significant evolution with time and frequency, attributable to relative changes in the leading, central, and trailing emission components. The magnetar generally spins down with an average rate of $-1.65(4)\times10^{-11}\,\mathrm{s^{-2}}$, with the magnitude of the spin-down rate first increasing and then decreasing over the monitoring span. The overall flux density shows an initial rise followed by a decline, with sporadic variability superimposed; the variations at 2.25 and 8.60~GHz are broadly consistent across most epochs. The two-point spectral index increased from $\alpha\simeq-1.8$ shortly after the outburst to near-zero values around MJD~59030--59050, followed by repeated decreases and increases. Despite this variability, 41 of the 54 two-point spectral-index measurements in the later campaign have $\alpha>-1$. Three component-dependent types of mode changing are identified: Type~I, involving primarily intensity changes in the central component; Type~II, involving central and trailing components; and Type~III, involving leading and central components. All three classes include events showing a corresponding component response in the other observing band, with event-dependent response strengths and directions. Type~I variations are generally more pronounced at 8.60~GHz than at 2.25~GHz.
\end{abstract}

\begin{keywords}
stars: magnetars -- stars: neutron -- pulsars: individual: Swift~J1818.0$-$1607
\end{keywords}

\section{Introduction}

Magnetars are isolated neutron stars with typically inferred surface dipolar magnetic fields of approximately $10^{13}$--$10^{15}$~G, whose radiative activity is powered primarily by the evolution and dissipation of magnetic energy \citep{dt1992,kb2017}. By early 2026, the known population comprised 32 confirmed or candidate magnetar systems\footnote{\url{https://www.physics.mcgill.ca/~pulsar/magnetar/main.html}} \citep{ok2014,cbi2021,enh2021}; periodic radio pulsations had been detected from only six of them \citep{bwp2023,fbr2024}. Unlike the comparatively stable emission of most ordinary pulsars, radio magnetars often show substantial changes in pulse profile, flux density, and radio spectrum \citep{crj2007,ccr2007,ags2012,tek2015,ljs2021}. Their radio activity has also been observed in connection with X-ray evolution, while changes in radio emission may accompany variations in rotational torque \citep{ccr2007,crp2007,rsl2022,bwp2023}. Radio observations therefore probe not only the coherent emission itself but also rotational evolution through timing measurements, making them particularly valuable for investigating the evolving magnetosphere and its response to underlying magnetic and crustal activity \citep{pbh2012,pbh2013,kb2017}.

Mode changing refers to rapid switching between two or more quasi-stable emission states with distinguishable mean pulse profiles \citep{b1970,l1971,bms1982,wmj2007}. In some pulsars, profile changes are accompanied by changes in spin-down rate, supporting an interpretation in terms of transitions between metastable magnetospheric configurations \citep{lhk2010}. Such switching is observationally distinct from the gradual profile and flux evolution seen between magnetar observing epochs, because mode changing involves recurrent transitions between quasi-stable emission states within an individual observing session, whereas the latter describes changes in the prevailing emission properties between observing epochs and does not necessarily involve discrete or recurrent states. Simultaneous multifrequency observations allow the transition times and mode-resolved pulse profiles to be compared directly between frequency bands. A corresponding profile-state transition in both bands provides evidence that the underlying switch is broadband, while differences in component intensity or flux contrast reveal a frequency-dependent radiative response \citep{ysm2018}.

Swift~J1818.0$-$1607 was discovered following a high-energy outburst detected by the \textit{Swift} Burst Alert Telescope on 2020 March 12 \citep{egk2020}. Follow-up observations established a spin period of approximately 1.36~s and detected periodic radio emission \citep{esp2020}. Early wide-band spectropolarimetry found a steep radio spectrum and strong linear polarization \citep{lsjb2020}. High-cadence monitoring found a relatively narrow and simple pulse profile and large spin-down variations \citep{ccc2020}, while later wide-band observations revealed profile and polarization evolution and short-time-scale emission-state switching \citep{ljs2021}. Longer-term monitoring further revealed systematic changes in both the spin-down behaviour and persistent pulse-profile states over time-scales of months \citep{rsl2022,fbr2024}. More recent observations showed a gigahertz-peaked broadband spectrum with component-dependent spectral properties \citep{lbl2025}, while later 2.0~GHz monitoring during a period of reduced activity found a comparatively stable narrow profile and a smaller magnitude of the spin-down rate \citep{afg2025}. The short-time-scale events include two distinct forms reported at 0.7--4.0~GHz, bright--quiet switching detected at three epochs in our earlier simultaneous 2.25/8.60~GHz campaign, and switching between two profile states observed simultaneously at 2.2 and 8.4~GHz \citep{ljs2021,hys2021,bwp2023}. Mode-changing or discrete radio-emission state transitions have also been reported in other radio magnetars, including 1E~1547.0$-$5408 and PSR~J1745$-$2900 \citep{crh2007,hgr2008,ywm2018}. Nevertheless, the repeated occurrence of different forms of short-time-scale emission-state switching in Swift~J1818.0$-$1607 at different stages of its post-outburst evolution makes it a particularly valuable source for investigating the pulse-component and frequency dependence of magnetar emission-state changes.

Swift~J1818.0$-$1607 is therefore a source for which long-term simultaneous multifrequency monitoring is particularly valuable. Its pulse profile, flux density, spectrum, spin-down behaviour, and short-time-scale emission states all evolve rapidly after outburst, so non-simultaneous or sparsely sampled observations can make it difficult to separate intrinsic frequency-dependent behaviour from temporal evolution. Our earlier simultaneous 2.25/8.60~GHz campaign with the Shanghai Tian Ma Radio Telescope (TMRT) followed Swift~J1818.0$-$1607 from MJD~58936 to 59092, capturing the early radio-active phase after the 2020 outburst \citep{hys2021}. During that interval, the source brightened at both frequencies, its two-point spectral index increased, its spin-down behaviour remained irregular, and bright--quiet switching was detected on MJD~59039, 59050, and 59051. These results raised several questions about the subsequent evolution: whether the early radio brightening and increase in the two-point spectral index persisted or reversed, whether the long-term spin-down evolution could still be represented by a single average slope, and whether the previously detected bright--quiet switching recurred or formed part of a broader family of short-time-scale state changes.

Here we present 65 additional simultaneous TMRT observations at 2.25 and 8.60~GHz from MJD~59104 to 59427, extending the same dual-frequency monitoring programme by a further 323~d. This data set connects the early radio-active phase reported in our previous work to the later post-outburst evolution, allowing us to examine the long-term evolution of the pulse profile, timing parameters, flux density, and two-point spectral index, and to compare later mode-changing events within a uniform simultaneous observing setup. Section~\ref{sec:obs} describes the observations and data reduction. Section~\ref{sec:results} presents the long-term evolution and identifies three phenomenological classes of short-time-scale mode changing. Section~\ref{sec:discussion} compares these results with previous studies and discusses their possible magnetospheric origin, and Section~\ref{sec:conclusions} summarizes our conclusions.

\section{Observations and Data Reduction} \label{sec:obs}

Following our earlier simultaneous 2.25/8.60~GHz TMRT campaign from MJD~58936 to 59092 \citep{hys2021}, we carried out 65 additional dual-frequency observations of Swift~J1818.0$-$1607 from MJD~59104 to 59427, spanning 323~d. The observing MJD and integration time, $T_{\rm obs}$, are listed in the first two columns of each half of Table~\ref{tab:parameters}; the integration times range from approximately 25 to 360~min. The simultaneous coverage and generally long observing durations provide a uniform data set for examining both long-term evolution and minute-time-scale variability.

The TMRT is a fully steerable 65~m Cassegrain antenna located in Shanghai \citep{wzy2016}. Its cryogenically cooled dual-polarization S/X-band receiver provides effective frequency coverage of 2.20--2.30~GHz and 8.20--9.00~GHz, respectively. The capability for simultaneous pulsar observations at 2.25 and 8.60~GHz with the TMRT was demonstrated as early as 2018 \citep{ysm2018}. The dual-frequency signals were sampled and processed simultaneously with the Digital Backend System (DIBAS), which consists of three parallel ROACH-2/ADC processing units \citep{ysm2018}. To reduce intra-channel dispersion smearing and facilitate frequency-domain radio-frequency interference (RFI) excision, the observing bands were divided into channels of approximately 1~MHz at 2.25~GHz and 2~MHz at 8.60~GHz.

We used the incoherent-dedispersion and online-folding mode implemented in DIBAS using functionalities inherited from the Green Bank Ultimate Pulsar Processing Instrument \citep{drd08}. The data were dedispersed at a dispersion measure of 701~pc~cm$^{-3}$ and folded online using predictive timing files generated with \texttt{TEMPO}\footnote{\url{https://tempo.sourceforge.net/}}. Each rotation period was divided into 1024 phase bins, and the data were recorded as 30~s subintegrations in the standard 8-bit pulsar-specific format based on the Flexible Image Transport System (FITS), known as PSRFITS \citep{hsm2004}. The folded data were processed with \texttt{PSRCHIVE}\footnote{\url{https://psrchive.sourceforge.net/}} \citep{hsm2004,sdo2012}. Each observation was inspected in the time--frequency and pulse-phase domains, and corrupted frequency channels or subintegrations were removed interactively using \texttt{pazi}. The two polarizations were then summed to form total-intensity profiles, and the data were averaged in time and frequency with \texttt{pam} as required for the subsequent analyses.

As no real-time flux calibration system was available during most observations, we estimated the flux densities from the observed signal-to-noise ratio (S/N) using the radiometer equation and the TMRT system equivalent flux densities (SEFDs) \citep{lrf11,zwy2017}. We adopted an instrumental correction factor of 1.5 and SEFDs of approximately 46 and 48~Jy at 2.25 and 8.60~GHz, respectively \citep{wzy2016,ysm2018}. The effective integration time and bandwidth after RFI excision were used for each observation. The mean flux density was obtained by integrating the calibrated profile over the on-pulse phase bins and dividing by the total number of phase bins; its uncertainty was estimated from the off-pulse root-mean-square (RMS) noise divided by the square root of the total number of phase bins \citep{zwy2017,hys2026}.

For epochs with significant detections in both bands, we calculated the two-point spectral index directly from the band-averaged flux densities at 2.25 and 8.60~GHz as
\begin{equation}
\alpha =
\frac{\ln(S_{8.60}/S_{2.25})}
{\ln(8.60/2.25)}.
\end{equation}
Thus, the reported values are two-point spectral indices defined by the two TMRT band-averaged flux-density measurements, rather than independent in-band spectral-index measurements within either receiver band.

For the mode-resolved in-band spectra of MJD~59105, we used the 30~s subintegrations assigned to State~A or B by the component-flux hidden Markov model (HMM) described in Section~\ref{sec:mode_changing} and valid in both observing bands. We measured the baseline-subtracted flux density of the central component in each frequency channel. Channel-dependent calibration factors were derived from on-source/off-source (ON/OFF) observations of 3C286 and 3C295 using their standard flux-density spectra \citep{perley2017} and applied to the corresponding target channels. After excluding RFI-affected and invalid channels, we combined the calibrated measurements with inverse-variance weights over 24 native channels per spectral point at 2.25~GHz and 16 at 8.60~GHz. We obtained the in-band spectral indices and their formal uncertainties from weighted linear fits in logarithmic frequency--flux space, separately within each band and state.

\begin{table*}
	\centering
	\caption{Observing and radio-emission parameters of Swift~J1818.0$-$1607.}
	\label{tab:parameters}
	\footnotesize
	\setlength{\tabcolsep}{0pt}
	\renewcommand{\arraystretch}{0.92}
	\begin{tabular*}{\textwidth}{@{\extracolsep{\fill}}lcccc@{\hspace{14pt}}lcccc@{}}
		\hline
		\textbf{MJD} & $T_{\rm obs}$ & $W_{50}$ & $S_{\rm 2.25}/S_{\rm 8.60}$ & $\alpha$ &
		\textbf{MJD} & $T_{\rm obs}$ & $W_{50}$ & $S_{\rm 2.25}/S_{\rm 8.60}$ & $\alpha$ \\
		\multicolumn{1}{c}{--} & (min) & (deg) & (mJy) & -- & \multicolumn{1}{c}{--} & (min) & (deg) & (mJy) & -- \\
		\noalign{\vskip 2pt}
		\cline{1-5}\cline{6-10}
		\noalign{\vskip 2pt}
59104 & 83.2 & 5.9(2)/6.12(5) & 1.6(1)/1.69(4) & 0.03(5) & 59298 & 44.1 & 9.7(2)/-- & 3.6(1)/-- & -- \\
		59105 & 83.2 & 6.1(2)/5.29(4) & 1.7(1)/2.02(4) & 0.13(5) & 59299 & 119.8 & 9.3(1)/-- & 3.28(8)/-- & -- \\
		59106 & 83.2 & 6.1(2)/5.52(4) & 1.5(1)/1.73(4) & 0.12(5) & 59301 & 178.4 & 10.03(9)/-- & 3.82(7)/-- & -- \\
		59109 & 59.6 & 6.1(3)/-- & 1.5(1)/-- & -- & 59302 & 119.8 & 10.8(2)/-- & 3.44(8)/-- & -- \\
		59110 & 59.6 & 6.5(2)/5.49(5) & 2.1(1)/1.59(4) & $-$0.19(5) & 59304 & 179.4 & 10.3(1)/13.4(4) & 3.34(7)/1.19(2) & $-$0.77(2) \\
		59111 & 88.2 & 7.5(2)/5.51(4) & 2.2(1)/1.76(3) & $-$0.16(4) & 59305 & 179.4 & 9.8(1)/14.5(3) & 2.95(7)/0.89(4) & $-$0.89(4) \\
		59112 & 188.9 & --/5.48(2) & --/1.90(3) & -- & 59306 & 112.2 & 9.8(2)/12.1(4) & 2.80(9)/1.15(3) & $-$0.66(3) \\
		59113 & 83.2 & 7.3(1)/5.26(3) & 3.0(1)/3.16(4) & 0.04(3) & 59314 & 179.4 & 9.48(9)/-- & 2.66(7)/-- & -- \\
		59115 & 92.2 & 7.5(2)/5.59(6) & 2.45(9)/1.25(3) & $-$0.50(4) & 59315 & 59.6 & 10.5(1)/-- & 3.8(1)/-- & -- \\
		59117 & 99.2 & 7.1(1)/5.41(2) & 2.54(9)/2.25(3) & $-$0.09(3) & 59316 & 218.5 & 10.10(9)/9.9(2) & 3.15(6)/0.96(5) & $-$0.89(4) \\
		59123 & 130.3 & 6.34(7)/5.26(1) & 2.69(8)/3.61(3) & 0.22(2) & 59319 & 178.9 & 10.1(1)/10.5(2) & 3.10(7)/0.63(2) & $-$1.19(3) \\
		59125 & 124.3 & 7.5(2)/5.56(7) & 1.63(8)/0.98(5) & $-$0.38(5) & 59321 & 74.7 & 9.8(1)/10.5(3) & 3.1(1)/0.64(4) & $-$1.17(5) \\
		59141 & 83.7 & 6.4(1)/5.67(5) & 3.0(1)/1.55(4) & $-$0.49(3) & 59329 & 59.6 & 9.9(2)/13.8(7) & 2.4(1)/0.82(5) & $-$0.79(6) \\
		59149 & 61.1 & 7.6(1)/-- & 4.1(1)/-- & -- & 59330 & 59.6 & 9.7(2)/12.0(3) & 2.7(1)/0.75(5) & $-$0.96(6) \\
		59153 & 76.7 & 7.00(6)/5.87(5) & 4.8(1)/2.76(4) & $-$0.42(2) & 59337 & 179.4 & 10.02(9)/16.7(4) & 2.90(7)/1.16(3) & $-$0.68(3) \\
		59154 & 89.7 & 6.90(5)/6.09(5) & 7.1(1)/2.97(3) & $-$0.65(1) & 59338 & 179.4 & 10.2(1)/17.3(1) & 2.76(7)/1.40(2) & $-$0.51(2) \\
		59157 & 123.8 & 6.48(3)/5.63(3) & 5.56(8)/1.75(3) & $-$0.86(2) & 59339 & 179.4 & 10.61(9)/16.2(1) & 3.36(7)/1.90(3) & $-$0.42(2) \\
		59159 & 83.7 & 6.48(5)/5.70(3) & 5.7(1)/3.45(4) & $-$0.38(2) & 59381 & 59.6 & 11.8(2)/14.1(5) & 2.7(1)/0.57(4) & $-$1.15(6) \\
		59168 & 59.6 & 6.40(6)/5.6(2) & 4.9(1)/1.09(4) & $-$1.13(3) & 59388 & 39.6 & 12.0(4)/18(2) & 1.5(1)/0.31(6) & $-$1.2(2) \\
		59208 & 264.6 & 7.45(4)/6.48(3) & 7.11(6)/3.70(2) & $-$0.488(7) & 59390 & 59.6 & 11.6(3)/14.6(8) & 2.7(1)/0.22(4) & $-$1.9(1) \\
		59212 & 179.9 & 7.40(6)/-- & 5.19(7)/-- & -- & 59391 & 59.6 & 11.3(2)/14.0(6) & 2.8(1)/0.37(5) & $-$1.5(1) \\
		59213 & 119.8 & 7.9(1)/6.5(1) & 3.80(9)/1.72(4) & $-$0.59(2) & 59392 & 59.6 & 18.1(3)/17.3(8) & 3.2(1)/0.36(4) & $-$1.63(9) \\
		59215 & 339.7 & 7.54(4)/6.67(2) & 5.55(5)/3.23(2) & $-$0.405(8) & 59393 & 59.6 & 11.7(2)/18.9(6) & 2.4(1)/0.50(4) & $-$1.18(7) \\
		59217 & 299.6 & 7.97(6)/7.00(3) & 5.00(5)/2.64(2) & $-$0.477(9) & 59394 & 59.6 & 12.9(2)/17.2(5) & 3.4(1)/0.51(4) & $-$1.42(7) \\
		59218 & 359.3 & 8.39(7)/6.72(4) & 4.92(5)/2.22(2) & $-$0.59(1) & 59395 & 59.6 & 12.7(2)/16.4(6) & 2.7(1)/0.44(4) & $-$1.33(8) \\
		59221 & 279.6 & 7.66(7)/6.68(3) & 3.42(5)/2.38(2) & $-$0.27(1) & 59396 & 119.3 & 12.4(2)/17.9(4) & 2.13(8)/0.89(5) & $-$0.65(5) \\
		59245 & 124.3 & --/7.66(7) & --/1.64(3) & -- & 59397 & 59.6 & 13.9(4)/17.1(2) & 2.5(1)/1.69(4) & $-$0.29(4) \\
		59277 & 24.8 & 8.3(1)/8.66(8) & 5.0(2)/1.99(3) & $-$0.69(3) & 59399 & 79.7 & 14.7(5)/19.2(1) & 2.7(1)/1.88(4) & $-$0.27(3) \\
		59283 & 149.3 & 8.97(6)/10.2(3) & 4.78(7)/0.56(5) & $-$1.60(7) & 59417 & 29.6 & 13.2(4)/12(1) & 1.5(2)/0.94(6) & $-$0.36(9) \\
		59287 & 88.7 & 9.21(8)/9.7(1) & 5.5(1)/1.95(3) & $-$0.78(2) & 59418 & 59.6 & 12.2(2)/12.5(3) & 1.6(1)/0.90(7) & $-$0.42(8) \\
		59290 & 35.1 & 10.5(2)/10.2(2) & 5.4(2)/1.92(4) & $-$0.77(3) & 59426 & 119.3 & 15.9(3)/13.6(3) & 1.21(8)/1.04(3) & $-$0.11(6) \\
		59293 & 59.1 & 9.1(2)/11.3(8) & 3.1(1)/0.74(5) & $-$1.07(6) & 59427 & 119.3 & 12.2(3)/13.8(2) & 0.82(8)/0.76(3) & $-$0.05(8) \\
		59295 & 39.6 & 9.8(2)/9.3(3) & 3.7(1)/1.23(6) & $-$0.82(5) & \multicolumn{1}{c}{--} & -- & -- & -- & -- \\
		\hline
	\end{tabular*}
	\renewcommand{\arraystretch}{1}
	\begin{minipage}{\textwidth}
		\footnotesize
		\emph{Notes.} $T_{\rm obs}$ is the integration time. The $W_{50}$ and flux-density columns are listed as 2.25/8.60~GHz values. $W_{50}$ is measured from the total one- to three-Gaussian fitted profile. Parentheses give $1\sigma$ uncertainties in the last quoted digits; the $W_{50}$ errors are formal fit-covariance uncertainties. The spectral index $\alpha$ is the two-point index calculated from the 2.25 and 8.60~GHz mean flux densities.
	\end{minipage}
\end{table*}

\section{Results}
\label{sec:results}

We examine the long-term evolution of the pulse-profile morphology, rotational properties, flux density, and radio spectrum before turning to the short-time-scale mode-changing behaviour detected within individual observations.

\subsection{Pulse profile evolution}

\begin{figure*}
	\centering
	 \includegraphics[width=0.90\textwidth]{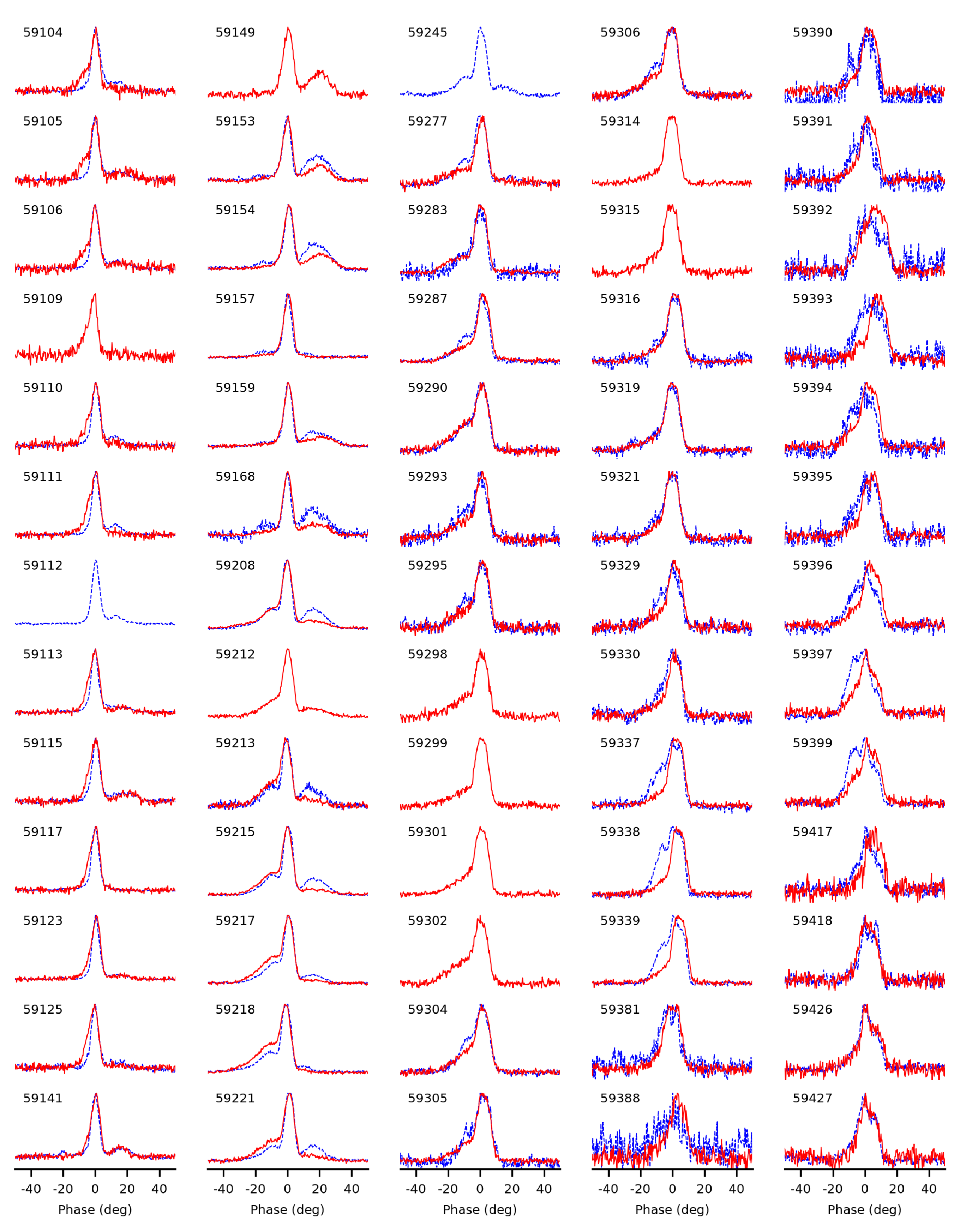}
	\caption{
		Integrated pulse profiles of Swift~J1818.0$-$1607 at 2.25 and 8.60~GHz over the interval MJD~59104--59427. The red solid and blue dashed lines represent the 2.25 and 8.60~GHz profiles, respectively. For epochs with detections in both bands, both profiles are normalized by their peak intensities, the peak of the higher-S/N 8.60~GHz profile is placed at phase $0\degr$, and the simultaneous 2.25~GHz profile is aligned relative to it using the timing solution. For epochs with a detection in only one band, the peak of the detected profile is placed at phase $0\degr$.
	}
	\label{fig:profiles}
	\par\smallskip
	\begin{minipage}{\linewidth}\small\raggedright
	\textbf{Alt text:} A chronological grid of aligned pulse profiles at 2.25 and 8.60~GHz from MJD~59104 to 59427. The profile changes from a mostly narrow central peak at early epochs to profiles with intermittent trailing emission and, at later epochs, broader asymmetric emission with a stronger leading side. The two frequencies generally align in phase but differ in their relative component strengths.
	\end{minipage}
\end{figure*}

The integrated pulse profiles at 2.25 and 8.60~GHz from MJD~59104 to 59427 are shown in Fig.~\ref{fig:profiles}. The original S- and X-band data were phase-aligned using the timing solution, allowing the profiles to be compared across epochs. Fig.~\ref{fig:profiles} shows clear long-term evolution: the early profiles are dominated by a relatively narrow central peak, trailing-side emission subsequently strengthens, leading-side emission becomes more prominent, and the later profiles become broader and more asymmetric. Visual inspection of profiles formed from shorter segments within individual observations indicates that the morphology is relatively stable on short time-scales in most observations. A subset instead shows switching between quasi-stable profiles; these mode-changing events are discussed below.

To quantify these changes, we fitted the 119 integrated profiles with multicomponent Gaussian models containing one to three terms per profile. We refer to the fitted terms according to their pulse-longitude positions as leading (L), central (C), and trailing (T) components. The longitude intervals used to measure the corresponding emission are $-38.67\degr \leq \phi < -5.63\degr$, $-5.63\degr \leq \phi < 8.79\degr$, and $8.79\degr \leq \phi \leq 36.21\degr$, respectively. For each interval, we calculate the baseline-subtracted integrated intensity and estimate its uncertainty from the scatter of integrations over off-pulse intervals of the same width. We regard emission in an interval as detected when its integrated signal-to-noise ratio is at least 3. Central emission is detected in every profile, whereas emission in the leading and trailing regions varies with epoch and frequency. The number of Gaussian terms used at each epoch and frequency, together with their amplitudes, centroid longitudes, full widths at half maximum (FWHMs) and component separations, is reported in Table~S1 of the Supplementary Material.

Based on the prevailing profile morphology, we describe the long-term evolution using four broad intervals. During MJD~59104--59125, the central component dominates and the profiles remain relatively compact, although weaker outer emission is already visible. Leading-side emission is generally more prominent at 2.25~GHz, whereas trailing-side emission is more readily discerned at 8.60~GHz. During MJD~59141--59168, the trailing component strengthens and extends the profile towards later pulse longitudes; its relative strength varies with epoch and frequency. During MJD~59208--59245, both leading and trailing emission contribute to broader, more complex multicomponent profiles. The overall morphology remains broadly similar between the two bands, although the relative strengths of the outer components differ.

During MJD~59277--59427, the profiles are generally broader, mainly owing to the extension and enhancement of leading-side emission rather than uniform broadening of the central component. The frequency dependence is clearest around MJD~59337--59339, when the leading-side excess is much more pronounced at 8.60~GHz than at 2.25~GHz. This morphology is consistent with later 2.0~GHz observations in which the asymmetric profile was described by a narrow main component and a broader precursor \citep{afg2025}. As leading-side emission strengthens, the trailing component tends to weaken or disappear. Broad profiles persist after MJD~59381, while the relative contributions of leading and central emission continue to vary with epoch and frequency. The four intervals describe the prevailing morphology at the sampled epochs. Their boundaries lie between the observations on either side of a noticeable morphological change; the intervening gaps prevent a precise transition time from being assigned. Representative profiles from the four broad intervals are shown in Fig.~S1 of the Supplementary Material.

We use the full width at half maximum, $W_{50}$, as the primary indicator of profile width. For each profile, we sum the Gaussian terms to obtain the fitted total profile, identify the half-maximum level relative to its peak, and determine the outermost half-maximum crossing longitudes by interpolation. Their separation gives $W_{50}$; the quoted uncertainties are formal errors propagated from the Gaussian-fit covariance. The measurements are listed in Table~\ref{tab:parameters} and plotted in Fig.~\ref{fig:evolution}(b). The variable leading and trailing emission makes low-intensity widths such as $W_{10}$ sensitive to whether weak outer emission is included. By comparison, $W_{50}$ mainly traces the central peak at early epochs while responding when an outer component becomes sufficiently strong. The profiles are relatively narrow at the start of the campaign, with slightly larger $W_{50}$ at 2.25~GHz than at 8.60~GHz. Their widths subsequently increase overall, and the 8.60~GHz profiles are broader at many later epochs. The difference is particularly clear around MJD~59337--59339, consistent with the stronger high-frequency leading emission at those epochs. Once the leading component reaches the half-maximum level of the total profile, it can substantially increase $W_{50}$. Thus, the long-term width evolution reflects changes in the multicomponent profile rather than solely an intrinsic broadening of the central peak.

\subsection{Timing evolution}

\begin{figure*}
	\centering
	 \includegraphics[width=0.93\textwidth]{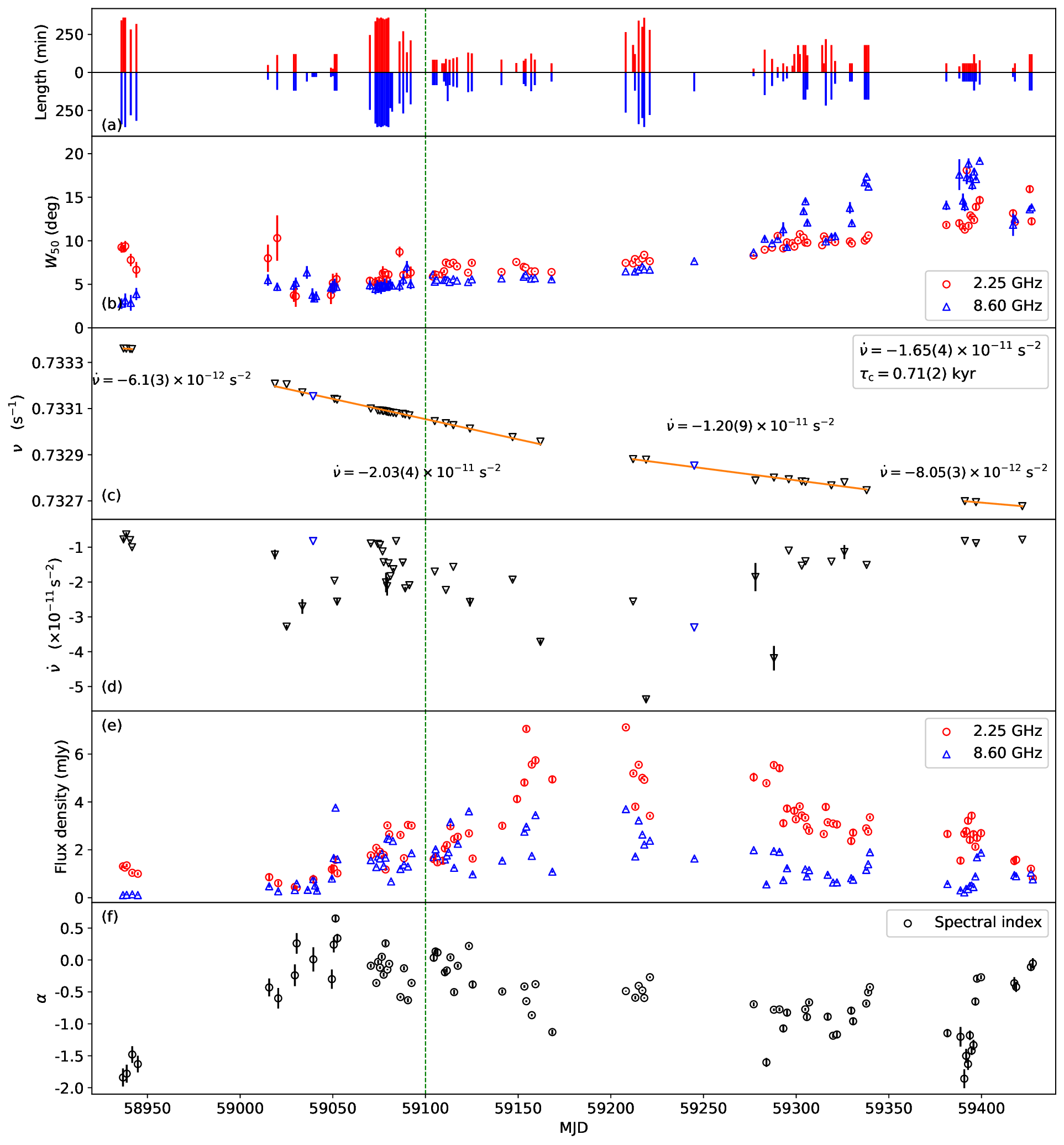}
	\caption{
		Long-term evolution of the rotational and radiative properties of Swift~J1818.0$-$1607. Panel (a): integration times of the 2.25~GHz (red upward bars) and 8.60~GHz (blue downward bars) observations used in the timing analysis. Panel (b): evolution of $W_{50}$ at 2.25 and 8.60~GHz; the new-campaign values are measured from the fitted total one- to three-Gaussian profiles, while the earlier measurements are retained from our previous work. Panel (c): spin-frequency measurements and linear fits over four broad time intervals. Panel (d): spin-down rates obtained independently from the individual timing segments. In panels (c) and (d), blue symbols mark timing measurements derived using only 8.60~GHz data. Panel (e): flux density evolution at 2.25 and 8.60~GHz. Panel (f): evolution of the 2.25--8.60~GHz two-point spectral index. The green dashed line separates the previous observing campaign from the new observations presented in this work.
	}
	\label{fig:evolution}
	\par\smallskip
	\begin{minipage}{\linewidth}\small\raggedright
	\textbf{Alt text:} Six vertically stacked time-series panels spanning MJD~58936--59427. The pulse width generally increases at late epochs, the spin frequency declines through four fitted intervals, and the spin-down rate varies between timing segments. The flux density rises and then declines with superposed variability, while the two-point spectral index varies repeatedly. A vertical dashed line separates the previous and new observing campaigns.
	\end{minipage}
\end{figure*}

The strong rotational instability and evolving pulse profiles of radio magnetars can prevent the construction of a single phase-coherent timing solution over an extended interval. We therefore adopted a segment-based timing procedure similar to that used in previous studies of the radio magnetar XTE~J1810$-$197 \citep{ccr2007,crd2022}. The data were divided into intervals within which the pulse-profile morphology remained sufficiently stable. Each segment was extended for as long as possible, until a significant profile change occurred or the relative uncertainty in the fitted spin-down rate exceeded 5\%. The observation that first caused the relative uncertainty to exceed 5 per cent was retained in the segment, so the final timing solution can have a relative uncertainty above this threshold. Frequency-specific standard templates were constructed from integrated profiles with high signal-to-noise ratios using \texttt{paas} in \texttt{PSRCHIVE}, and times of arrival (ToAs) were measured by cross-correlating each profile with the corresponding template using \texttt{pat} \citep{hsm2004,sdo2012}. ToAs from both observing bands were combined for most segments. For a small number of observations made while the 2.25~GHz receiver was under maintenance, only the 8.60~GHz data were used; the resulting timing measurements are marked in blue in Fig.~\ref{fig:evolution}(c) and (d). The integration times of the data used for timing are shown in Fig.~\ref{fig:evolution}(a). For each segment, we fitted the spin frequency $\nu$ and its first derivative $\dot{\nu}$ at the segment reference epoch using \texttt{TEMPO2} \citep{hem2006,ehm2006}. Phase coherence was maintained within each individual timing segment. The corresponding timing parameters and residual statistics are provided in Table~S2 of the Supplementary Material.

The spin-frequency measurements are shown in Fig.~\ref{fig:evolution}(c). To provide a descriptive characterization of the long-term spin-frequency evolution, we fitted the measured $\nu$ values linearly within four broad time intervals. These intervals were adopted as a convenient parametrization of the changing slope of the long-term spin-frequency decline rather than as statistically distinct rotational stages. The fitted intervals span MJD~58937--58941, 59018--59162, 59212--59338, and 59391--59422, with the endpoints corresponding to the first and last measurements included in each interval rather than precisely determined transition epochs. The corresponding slopes give average spin-down rates of approximately $-6.1(3)\times10^{-12}$, $-2.03(4)\times10^{-11}$, $-1.20(9)\times10^{-11}$, and $-8.05(3)\times10^{-12}\,\mathrm{s^{-2}}$, respectively, where the uncertainties are the 1$\sigma$ errors from the linear fits. Thus, the magnitude of the long-term spin-down rate initially increased substantially and subsequently declined. A linear fit to all spin-frequency measurements over the full observing span gives an average $\dot{\nu}=-1.65(4)\times10^{-11}\,\mathrm{s^{-2}}$. Using the conventional definition $\tau_c=-\nu/(2\dot{\nu})$, we obtain an effective characteristic spin-down time-scale of $0.71(2)$~kyr for the average $\dot{\nu}$ over the analysed interval. Because the spin-down rate varies substantially with time, this value should not be interpreted as a reliable estimate of the magnetar's physical age.

The local spin-down rates fitted independently within the individual timing segments are shown in Fig.~\ref{fig:evolution}(d). These measurements display substantial segment-to-segment variations superposed on the broad evolution inferred from the spin-frequency fits. The interval with the strongest $\dot{\nu}$ variability broadly overlaps the period of complex multicomponent profiles and pronounced changes in the leading and trailing emission. At later epochs, the magnitude and scatter of $\dot{\nu}$ generally decrease while the profiles remain broad. Because the timing segments were defined partly according to profile stability, however, this apparent temporal correspondence is not fully independent. We therefore interpret it only as a possible qualitative association between the long-term profile and spin-down evolution, rather than as evidence for a direct correlation.

\subsection{Flux density and spectral evolution}

The mean flux densities of Swift~J1818.0$-$1607 at 2.25 and 8.60~GHz are listed in Table~\ref{tab:parameters} and plotted in Fig.~\ref{fig:evolution}(e). The measurements at 2.25~GHz are indicated by red open circles, while those at 8.60~GHz are shown as blue open triangles. The green dashed line separates our previous observations from the new observing campaign. Together, the two campaigns provide a more complete view of the radio evolution since the 2020 outburst. The source did not reach its maximum radio brightness immediately after the outburst. Instead, it went through an initial faint phase, followed by an overall brightening with substantial epoch-to-epoch fluctuations, and a later decline. During the earliest observations, the mean flux density was only about 1.0--1.4~mJy at 2.25~GHz and approximately 0.1~mJy at 8.60~GHz. After an early decrease to about 0.42~mJy at 2.25~GHz, the source brightened overall. In the new observing campaign, the 2.25~GHz flux density reached 7.11~mJy on MJD~59208, the highest value in the combined TMRT data set. At the same epoch, the 8.60~GHz flux density reached 3.70~mJy, which is the largest value in the new campaign and is comparable to the high values measured near the end of the earlier campaign.

Substantial fluctuations are superposed on this long-term evolution. During MJD~59141--59250, the flux densities typically range from about 3.0 to 7.1~mJy at 2.25~GHz and from about 1.1 to 3.7~mJy at 8.60~GHz. This interval broadly overlaps the period of complex multicomponent profiles and enhanced scatter in the segment-based spin-down measurements, again indicating similar stage-dependent behaviour among the radiative and rotational observables. The later decline is also not a smooth monotonic fading at either frequency. The weakening of the high-frequency emission is most evident during MJD~59381--59395, when the 8.60~GHz flux density dropped to about 0.2--0.6~mJy. It then temporarily recovered to about 1.7--1.9~mJy on MJD~59397--59399, and by MJD~59426--59427 the flux densities in the two bands had again become comparable. Thus, the late-time high-frequency fading is best described as a stage-dependent change rather than a persistent monotonic decline.

The two-point spectral indices calculated from the band-averaged 2.25 and 8.60~GHz flux densities, as described in Section~\ref{sec:obs}, are listed in Table~\ref{tab:parameters} and shown in Fig.~\ref{fig:evolution}(f). Shortly after the outburst, the two-point index was strongly negative, with $\alpha$ ranging from about $-1.84$ to $-1.48$ during MJD~58936--58944. It then increased rapidly, reaching values near zero or above zero at several epochs during MJD~59030--59051, with a maximum of about $+0.65$. At the beginning of the new campaign, the measurements during MJD~59104--59125 were mostly close to zero but still covered a range from $-0.50$ to $+0.22$. During MJD~59141--59250, most values were between about $-0.3$ and $-1.1$.

At later epochs, the two-point index continued to vary nonmonotonically. During MJD~59277--59330, the index generally decreased again, with $\alpha$ ranging from about $-0.7$ to $-1.6$. It then rose to between $-0.68$ and $-0.42$ during MJD~59337--59339. A second decrease occurred during MJD~59381--59395, reaching a minimum of about $-1.86$ on MJD~59390. This was followed by another rapid increase in the two-point index during MJD~59396--59427, with the final measurement reaching $\alpha\simeq-0.05$. The two-point index therefore shows repeated decreases and increases rather than a single monotonic trend.

The large epoch-to-epoch changes in both flux density and the two-point spectral index motivate a closer inspection of individual observations. In the following section, we examine whether short-time-scale switching between quasi-stable emission states contributes to the observed variability.

\subsection{Mode changing}
\label{sec:mode_changing}

Our long-term monitoring shows that the integrated pulse profile of Swift~J1818.0$-$1607 evolves substantially across the full observing span, while the profiles within each broad evolutionary interval retain a prevailing morphology (Fig.~\ref{fig:profiles}). This stage-dependent behaviour suggests that the post-outburst magnetosphere evolves on long time-scales while remaining comparatively stable over shorter intervals. Superposed on this evolving baseline, some individual observations show recurrent switching between two distinguishable, quasi-stable profile states. We refer to this short-time-scale, intra-observation behaviour as mode changing.

To identify these events, we examined the 30~s subintegrations of all 65 observations. After subtracting the off-pulse baseline and applying the flux-density calibration, we measured the integrated flux in the leading, central, and trailing pulse-phase regions of every valid subintegration. For each observation, we fitted the three component-flux series independently at 2.25 and 8.60~GHz with a two-state, full-covariance Gaussian HMM \citep{rabiner1989,jankowski2025hmm}, and compared it with a single-state vector autoregressive model of order one, VAR(1), representing continuous intensity variability. The HMM fit used 20 initializations. A statistically supported two-state candidate required a Bayesian information criterion (BIC) difference of $\Delta{\rm BIC}={\rm BIC}_{\rm VAR}-{\rm BIC}_{\rm HMM}\geq10$ \citep{schwarz1978,kass1995}, a run of at least three consecutive subintegrations in each state, and no pathological-fit or off-pulse quality flag. Each state had to occupy at least 10\% of valid subintegrations. Subintegrations with a maximum state posterior probability below 0.8 were left unassigned. Passing the screening in either band sufficed for confirmation. The final screening identified 15 confirmed mode-changing observations. MJD~59104 and 59217 did not pass the final screening but exhibit visually suggestive switching and are therefore identified separately as tentative events.

We classify the events phenomenologically by the pulse-phase region most strongly affected by the state transition. Type~I is dominated by the central component; Type~II involves the central and trailing components, with the trailing component providing the clearest state distinction; and Type~III involves the leading and central components, with enhanced leading-side emission and a broader profile. This classification describes the dominant change without requiring all other pulse phases to remain unchanged. For each event, we selected the band in which the switching is more conspicuous as the dominant band and used its HMM state sequence as the reference. The dominant band is 2.25~GHz for MJD~59337 and 8.60~GHz for the other events. We designate the reference-band state with the stronger type-defining component as State~A and the other as State~B. This convention fixes the state labels without requiring the corresponding component in the other band to change in the same direction.

Figures~\ref{fig:type1}--\ref{fig:type3} show MJD~59105, 59208, and 59399 as representative Type~I, II, and III events, respectively. Panels (a) show the 30~s subintegration intensity maps at both frequencies, the profiles averaged over the two states, and the standardized intensity time series of the relevant components,
\begin{equation}
    Z_0=\frac{I-\langle I\rangle}{\sigma_I},
\end{equation}
where $I$ is the component intensity in a subintegration and $\langle I\rangle$ and $\sigma_I$ are its mean and standard deviation during the observation. The state-averaged profiles use subintegrations that are valid and time-matched in both bands. The leading, central, and trailing component series, when displayed, are shown in orange, green, and purple. Panels (b) show eigenprofiles obtained from principal component analysis (PCA), the residual profiles relative to the observation-averaged profile, and the principal-component time series (PCTS) \citep{lin2018pca}. Individual principal components (PCs) are labelled PC1, PC2, and so on. The percentages beside the eigenprofiles give their fractions of the total profile variance. PCA describes the pattern of profile variability; the state assignments come from the component-flux HMM. Comparing the independently measured component-intensity and PCTS variations in the two bands gives a visual indication of whether the changes occur over corresponding intervals. Diagnostic plots for the remaining confirmed and tentative events are provided in Figs.~S2--S15 of the Supplementary Material.

When the response in the other band is weak, it may be difficult to assess from its intensity map alone. We therefore compare the mean integrated flux densities, $\bar S_A$ and $\bar S_B$, of the corresponding component in that band over the time-matched reference State~A and State~B subintegrations, using
\begin{equation}
    R_c=\frac{\bar S_A-\bar S_B}{(\bar S_A+\bar S_B)/2}\times100\%.
\end{equation}
For Types~I, II, and III, the response component is central, trailing, and leading, respectively. Positive $R_c$ means that the component in the other band is stronger during reference State~A; negative $R_c$ means that it is weaker. We also report a two-sided $p$ value for the difference between the two state-conditioned means, using heteroscedasticity- and autocorrelation-consistent errors \citep{neweywest1987} with a maximum lag of four subintegrations. This tests the other band's component response to the reference state, rather than an independently detected two-state structure in that band. The results for all 17 events are given in Table~\ref{tab:mode_response}.

\begin{table}
    \centering
    \caption{Cross-band component response for the 15 confirmed and two tentative mode-changing events.}
    \label{tab:mode_response}
    \footnotesize
    \begin{tabular*}{\columnwidth}{@{\extracolsep{\fill}}l c c r r}
        \hline
        Type & MJD & Dominant band & $R_c$ (\%) & Two-sided $p$ \\
        \hline
        I   & $59104^{\ast}$ & 8.60~GHz & $+39(6)$ & $1.21\times10^{-8}$ \\
        I   & 59105          & 8.60~GHz & $+18(6)$ & $0.00278$ \\
        I   & 59106          & 8.60~GHz & $+23(5)$ & $1.20\times10^{-5}$ \\
        I   & 59115          & 8.60~GHz & $-27(6)$ & $5.52\times10^{-6}$ \\
        I   & 59117          & 8.60~GHz & $-8(5)$ & $0.0823$ \\
        II  & 59208          & 8.60~GHz & $+149(7)$ & $1.59\times10^{-24}$ \\
        II  & 59215          & 8.60~GHz & $+100(10)$ & $3.32\times10^{-16}$ \\
        II  & $59217^{\ast}$ & 8.60~GHz & $+100(10)$ & $1.05\times10^{-4}$ \\
        II  & 59221          & 8.60~GHz & $+161(7)$ & $2.23\times10^{-17}$ \\
        III & 59337          & 2.25~GHz & $+100(20)$ & $5.97\times10^{-4}$ \\
        III & 59338          & 8.60~GHz & $+143(8)$ & $3.06\times10^{-16}$ \\
        III & 59339          & 8.60~GHz & $+119(9)$ & $1.33\times10^{-15}$ \\
        III & 59392          & 8.60~GHz & $+140(20)$ & $8.80\times10^{-4}$ \\
        III & 59393          & 8.60~GHz & $+170(10)$ & $0.0372$ \\
        III & 59394          & 8.60~GHz & $+160(10)$ & $0.00329$ \\
        III & 59397          & 8.60~GHz & $+191(8)$ & $9.89\times10^{-9}$ \\
        III & 59399          & 8.60~GHz & $+183(8)$ & $9.46\times10^{-11}$ \\
        \hline
    \end{tabular*}
    \par\smallskip
    \begin{minipage}{\columnwidth}\footnotesize
        \emph{Notes.} An asterisk marks a visually suggestive event that did not pass the final mode-changing screening. $R_c$ measures the response of the component in the other band to the dominant-band State~A/B classification. Parentheses give $1\sigma$ uncertainties in the last quoted digits, rounded to one significant digit, propagated from the heteroscedasticity- and autocorrelation-consistent covariance of the state-conditioned means (maximum lag: four subintegrations). The two-sided $p$ values test this conditional response and are not mode-detection probabilities.
    \end{minipage}
\end{table}

\subsubsection{Type~I mode changing}

\begin{figure*}
    \centering
    \includegraphics[width=\textwidth]{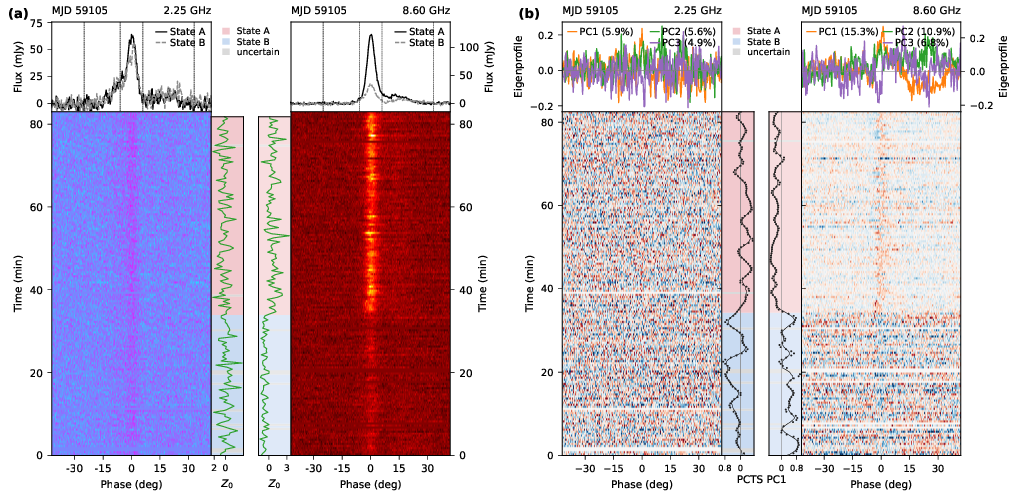}
    \caption{Representative Type~I mode-changing event on MJD~59105 at 2.25 and 8.60~GHz. Panel (a) shows the 30~s subintegration intensity maps, State~A/B mean profiles, and the $Z_0$ series of the central component (green). Panel (b) shows the eigenprofiles, residual-profile evolution, and central-dominated PCTS. Red and blue time-series backgrounds mark reference States~A and B, respectively; grey indicates uncertain or unmatched subintegrations. PCTS points are black. State-averaged profiles use only valid, time-matched subintegrations. The central-component flux contrast is much clearer at 8.60~GHz.}
    \label{fig:type1}
    \par\smallskip
    \begin{minipage}{\linewidth}\small\raggedright
    \textbf{Alt text:} Two side-by-side composite panels for MJD~59105. The 8.60~GHz time--phase map alternates between bright and faint central-component intervals, whereas the 2.25~GHz map shows a weaker contrast. Central-component intensity and PCA variations in the two bands occur over broadly corresponding intervals; red and blue shading show the reference State~A/B assignments.
    \end{minipage}
\end{figure*}

Type~I is confirmed on MJD~59105, 59106, 59115, and 59117. MJD~59104 fails the 10\% state-occupancy criterion (12/166 valid 8.60~GHz subintegrations, 7.2\%) but is retained as tentative on visual grounds. In the representative MJD~59105 observation, Fig.~\ref{fig:type1}(a) shows clear bright--faint central-component switching at 8.60~GHz; the corresponding total-flux contrast at 2.25~GHz is much weaker. In Fig.~\ref{fig:type1}(b), PC1 describes central-dominated profile variability in both bands, with PCTS variations occurring over broadly corresponding intervals. Conditional on the 8.60~GHz state assignment, the 2.25~GHz central component has $R_c=+18.4\%$ and a two-sided $p=0.00278$, quantifying the weaker low-frequency response.

Using the HMM-defined State~A/B subintegration groups, the 2.25--8.60~GHz two-point spectral indices on MJD~59105 are $+0.308\pm0.050$ and $-0.459\pm0.100$, respectively. They indicate different relative flux densities at the two observing frequencies. The low-frequency response does not have the same direction in every Type~I event: the central component increases in reference State~A on MJD~59106 but decreases on MJD~59115. The difference on MJD~59117 is weaker, with a two-sided $p=0.0823$ (Table~\ref{tab:mode_response}).

\subsubsection{Type~II mode changing}

\begin{figure*}
    \centering
    \includegraphics[width=\textwidth]{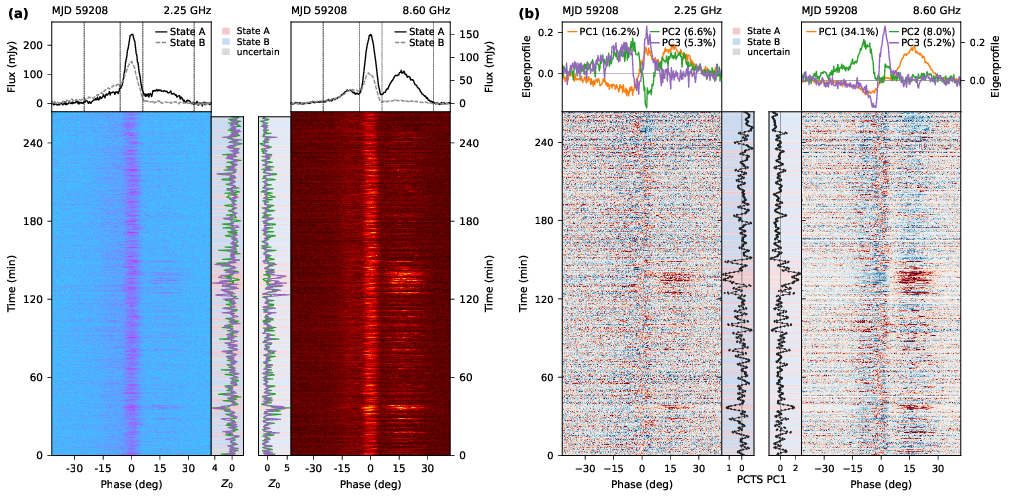}
    \caption{Representative Type~II mode-changing event on MJD~59208 at 2.25 and 8.60~GHz. Panel (a) shows the 30~s subintegration intensity maps, State~A/B mean profiles, and the $Z_0$ series of the central (green) and trailing (purple) components. Panel (b) shows the eigenprofiles, residual-profile evolution, and trailing-dominated PCTS. Red and blue time-series backgrounds mark reference States~A and B, respectively; grey indicates uncertain or unmatched subintegrations. PCTS points are black. State-averaged profiles use only valid, time-matched subintegrations. State~A has the stronger trailing component.}
    \label{fig:type2}
    \par\smallskip
    \begin{minipage}{\linewidth}\small\raggedright
    \textbf{Alt text:} Two side-by-side composite panels for MJD~59208. Both frequency maps show corresponding intervals of enhanced trailing emission and a changed central-to-trailing intensity ratio. Central and trailing intensity series and the trailing-dominated PCTS vary over similar intervals in the two bands. Red and blue backgrounds indicate reference States~A and B.
    \end{minipage}
\end{figure*}

Type~II is confirmed on MJD~59208, 59215, and 59221. On MJD~59217, the HMM did not significantly outperform continuous VAR(1) variability in either band ($\Delta{\rm BIC}<0$), but visual inspection motivated a tentative designation. These events involve changes in the relative strengths of the central and trailing components, with trailing enhancement providing the clearest distinction. In the representative MJD~59208 observation, Fig.~\ref{fig:type2}(a) shows recurrent intervals of enhanced trailing emission extending over several adjacent subintegrations. The trailing-component intensity variations in the two bands closely correspond in time, and the PCTS in Fig.~\ref{fig:type2}(b) also traces profile changes over similar intervals. With 8.60~GHz as the dominant band, the 2.25~GHz trailing component has $R_c=+149.1\%$ and a two-sided $p=1.59\times10^{-24}$. The other two confirmed Type~II events also show a positive trailing-component response at 2.25~GHz (Table~\ref{tab:mode_response}).

\subsubsection{Type~III mode changing}

\begin{figure*}
    \centering
    \includegraphics[width=\textwidth]{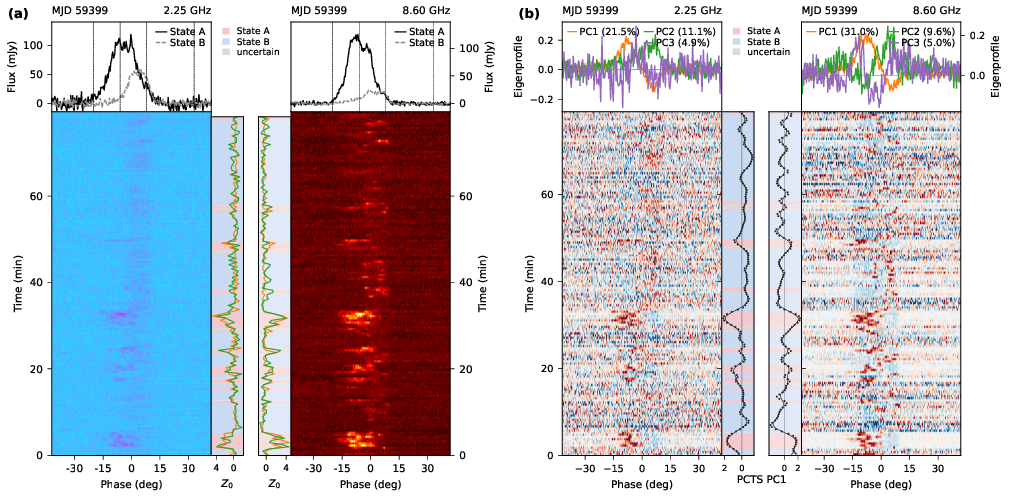}
    \caption{Representative Type~III mode-changing event on MJD~59399 at 2.25 and 8.60~GHz. Panel (a) shows the 30~s subintegration intensity maps, State~A/B mean profiles, and the $Z_0$ series of the leading (orange) and central (green) components. Panel (b) shows the eigenprofiles, residual-profile evolution, and leading-dominated PCTS. Red and blue time-series backgrounds mark reference States~A and B, respectively; grey indicates uncertain or unmatched subintegrations. PCTS points are black. State-averaged profiles use only valid, time-matched subintegrations. State~A has stronger leading-side emission and a broader profile.}
    \label{fig:type3}
    \par\smallskip
    \begin{minipage}{\linewidth}\small\raggedright
    \textbf{Alt text:} Two side-by-side composite panels for MJD~59399. Recurrent leading-to-central emission enhancements in both bands occur over corresponding intervals, and the State~A mean profile is broader and more asymmetric. The component-intensity and leading-dominated PCTS changes show good temporal correspondence across the two frequencies.
    \end{minipage}
\end{figure*}

Type~III events are confirmed on MJD~59337, 59338, 59339, 59392, 59393, 59394, 59397, and 59399. They involve the leading and central components; enhanced leading-side emission makes the State~A profile broader and more asymmetric. In the representative MJD~59399 observation, Fig.~\ref{fig:type3}(a) shows recurrent leading-side enhancements at both frequencies over corresponding intervals. The component-intensity and PCTS variations in the two bands also show good temporal correspondence (Fig.~\ref{fig:type3}(b)), with the leading-dominated eigenprofile labelled PC1 at both frequencies. With 8.60~GHz as the dominant band, the 2.25~GHz leading component has $R_c=+183.2\%$ and a two-sided $p=9.46\times10^{-11}$. All other Type~III events show a positive leading-component response in the corresponding other band; MJD~59337 uses 2.25~GHz as its dominant band (Table~\ref{tab:mode_response}).

The three classes are distinguished primarily by the pulse-phase region most affected by the transition: the central, trailing, or leading component, respectively. Both observing bands can show corresponding component-flux or profile variations, while the strength and direction of the flux response vary among events.

\section{Discussion}
\label{sec:discussion}

Our simultaneous dual-frequency monitoring extends the radio evolution of Swift~J1818.0$-$1607 to MJD~59427, revealing both stage-dependent long-term evolution and several forms of minute-time-scale state switching. We first compare these results with previous studies and then consider their implications for the magnetospheric evolution of the source.

\subsection{Long-term evolution in the context of previous studies}

The present observations extend and refine the early post-outburst picture established by our previous TMRT campaign. We first consider the pulse-profile evolution. In the earlier TMRT campaign, synchronized profile changes between 2.25 and 8.60~GHz were mainly associated with the trailing P3 component near MJD~59086--59092 \citep{hys2021}. In the present data, the outer components continue to dominate much of the profile variability, but the component responsible for the largest changes evolves with epoch. The initially compact profile develops stronger leading and trailing emission, and the late-time increase in $W_{50}$ is mainly driven by enhanced leading emission rather than by a simple broadening of the central component. On much longer time-scales, \citet{rsl2022} identified four radio emission modes at 1.4~GHz over their approximately one-year monitoring campaign, with the prevailing average profile remaining comparatively stable on time-scales of months. These long-lived profile states are distinct from the Type~I--III events analysed here, which involve transitions within individual observing sessions on time-scales of minutes. Nevertheless, both forms of variability show that the relative contributions of different pulse-longitude regions, including the leading, central, and trailing emission, change during the post-outburst evolution, although on very different time-scales. This component-dependent profile evolution is also consistent with wide-band Green Bank Telescope observations, which found a narrow central component flanked by leading and trailing components with stronger modulation and flatter spectra \citep{lbl2025}.

The timing evolution likewise extends the earlier picture. Our previous TMRT campaign followed Swift~J1818.0$-$1607 from MJD~58936 to 59092 and showed that its spin-frequency evolution was dominated by a long-term decrease with an average spin-down rate of approximately $-2.25\times10^{-11}\,\mathrm{s^{-2}}$ \citep{hys2021}. The extended timing baseline confirms the continued decrease of the spin frequency, but shows that the evolution cannot be represented by a single constant spin-down rate over the full radio-active interval. For a descriptive characterization of this changing slope of the long-term spin-frequency decline, we divide the spin-frequency measurements into four broad intervals with different average slopes. The average value obtained from the combined data set, $\dot{\nu}=-1.65(4)\times10^{-11}\,\mathrm{s^{-2}}$, corresponds to an effective characteristic spin-down time-scale of $0.71(2)$~kyr. This value applies to the radio-active phase and should not be regarded as a stable characteristic age, because the torque evolves strongly after the outburst. Published values inferred at different stages of the post-outburst evolution span a wide range, from approximately 0.27~kyr shortly after the outburst to about 0.5~kyr in subsequent early timing studies, $\sim0.86$~kyr from longer-term phase-coherent timing, and $\sim2.5$~kyr at later epochs \citep{ccc2020,hys2021,rsl2022,afg2025}. Our value of 0.71(2)~kyr lies within this evolving range, further illustrating the strong dependence of the inferred characteristic spin-down time-scale on the timing interval and torque state. This picture is broadly consistent with the coherent 1.4~GHz timing analysis of \citet{rsl2022}, who found that the spin-frequency derivative varied systematically around a mean value of about $-1.37\times10^{-11}\,\mathrm{s^{-2}}$ and that the modulation amplitude decreased with time. The difference in emphasis is partly methodological: their Gaussian-process analysis traces a smooth evolution of the spin-down modulation, whereas our segment-based analysis summarizes the long-term spin-frequency evolution through local timing solutions and broad descriptive intervals in the simultaneous dual-frequency data.

The flux-density evolution also extends, but does not simply continue, the early trends. During the previous TMRT campaign, the source brightened at both 2.25 and 8.60~GHz \citep{hys2021}. Early radio observations by Parkes and the Deep Space Network also showed substantial profile variability, initial steep spectra, subsequent radio brightening, and spectral flattening \citep{ljs2021,bwp2023}. Our later TMRT observations show that the brightening phase did not continue indefinitely: after reaching a maximum, the radio emission declined overall but retained substantial epoch-to-epoch fluctuations, including a temporary recovery of the 8.60~GHz emission near MJD~59397--59399. This nonmonotonic radio evolution is qualitatively consistent with the published 1.4~GHz light curve of \citet{rsl2022}, which likewise shows substantial post-outburst variability rather than a simple monotonic decay. A comparison of the digitized 1.4~GHz light curve with the TMRT 2.25 and 8.60~GHz measurements is presented in Fig.~S16 of the Supplementary Material. The Neutron star Interior Composition Explorer (NICER) measurements reported by \citet{rsl2022}, spanning approximately MJD~58927--59296 and therefore overlapping only the earlier part of our new TMRT campaign, showed an overall monotonic decline of the 0.5--12~keV X-ray flux. During the overlapping interval, the radio flux evolution does not simply track this X-ray decline, while our later radio measurements extend beyond the NICER coverage considered by \citet{rsl2022}.

Finally, the evolution of the two-point spectral index shows that the strongly negative values seen early in the monitoring were not representative of the later radio-active interval. Among the 54 epochs with usable detections in both bands, 41 have two-point spectral indices greater than $-1$, and 36 have indices greater than $-0.8$. These proportions characterize the dual-band detected sample rather than the fraction of monitoring time spent in a particular spectral state. The requirement of detections in both bands may bias the sample against steep-spectrum epochs with weak or undetected 8.60~GHz emission. The decline in 8.60~GHz flux density and decrease in the two-point index during MJD~59381--59395 were followed by a temporary flux recovery and rapid increase in the index, showing that the relative flux densities at our two frequencies remained highly variable. This behaviour is consistent with wide-band observations showing that the spectrum cannot be described by a single power law over the full radio band, including a spectral peak at $5.4(6)$~GHz and steep in-band spectra above 9~GHz \citep{lbl2025}. It also shows that a two-point spectral index measured between 2.25 and 8.60~GHz cannot fully characterize the broadband radio spectrum, because different profile components may contribute differently across frequency.

Taken together, the evolution of the profile, timing properties, flux density, and two-point spectral index indicates stage-dependent but non-identical changes in the radiative and rotational properties. The interval of higher radio brightness and more complex multicomponent profiles broadly overlaps the period in which the spin-down measurements show greater scatter, whereas both the radio flux and torque variability become less pronounced at late times. Although the sparse and segmented timing measurements do not warrant a formal correlation analysis here, the broad association suggests that the radio flux density, profile morphology, and braking torque respond to the same evolving post-outburst magnetosphere.

Comparable behaviour has been observed in other radio magnetars. XTE~J1810$-$197 has shown large changes in profile morphology, flux density, spectrum, and torque between different radio-active episodes \citep{ccr2007,hys2026}, while PSR~J1622$-$4950 and the Galactic-centre magnetar PSR~J1745$-$2900 have also exhibited profile reconfiguration and strong flux and spectral variability \citep{ags2012,tek2015}. In contrast to the highly stable average profiles of most ordinary pulsars \citep{hmt1975}, the observed changes in radio-magnetar emission are consistent with an evolving post-outburst magnetosphere. The particular value of Swift~J1818.0$-$1607 is that the combined TMRT campaigns provide a uniform simultaneous dual-frequency view of this evolution, showing broadly contemporaneous, but not necessarily directly correlated, changes in the pulse profile, spin-down behaviour, flux density, and two-point spectral index, with different amplitudes and characteristic time-scales.

\subsection{Comparison of the mode-changing behaviour}

Short-time-scale emission-state changes have been detected from Swift~J1818.0$-$1607 at several stages of its post-outburst evolution. Two forms of minute-time-scale switching were reported in two 0.7--4.0~GHz observations \citep{ljs2021}. In our earlier TMRT campaign, bright--quiet switching was detected on MJD~59039, 59050, and 59051, with the clearest event on MJD~59051 occurring simultaneously at 2.25 and 8.60~GHz \citep{hys2021}. The bright--quiet events reported in our earlier TMRT campaign show morphological similarities to the Type~I behaviour defined in the present work, particularly the central-component-dominated state changes and the stronger flux variations at 8.60~GHz. Because these earlier events were not analysed using the full classification procedure adopted here, we conservatively refer to them as likely Type~I analogues rather than as formal Type~I detections. Switching between two profile states was also observed simultaneously at 2.2 and 8.4~GHz on 2020 July 15 \citep{bwp2023}.

The four confirmed Type~I events on MJD~59105, 59106, 59115, and 59117 show that short-time-scale central-component changes similar to the earlier bright--quiet events were also present during the later radio-active phase. The 15 confirmed events reported in Section~\ref{sec:mode_changing} refer only to the present observing campaign; the three earlier TMRT epochs are described more cautiously as likely Type~I analogues because they were not analysed using the present classification procedure. Together, the four later Type~I events and three earlier likely analogues show that similar short-time-scale switching occurred at multiple TMRT epochs spanning MJD~59039--59117, rather than in a single isolated episode shortly after the outburst.

The earlier bright--quiet events and later Type~I events share the principal characteristic of short-time-scale central-component changes, with the bright--faint total-flux contrast generally more conspicuous at 8.60~GHz than at 2.25~GHz. The earlier observations showed changes in both bands, particularly on MJD~59051. In the present campaign, the component-flux HMM defines the reference states, while the 2.25~GHz central-component response to those states can be measured even when the total-flux contrast is weak, as on MJD~59105. The direction and strength of this low-frequency response vary among the later Type~I events. Thus, Type~I has a strongly frequency-dependent flux response.

The present observations further reveal two additional component-dependent forms of mode changing. Type~II involves changes spanning the central and trailing components, with enhancement of the trailing component providing the clearest distinction between the two states. Type~III instead involves the leading and central components and is characterized by enhanced leading-side emission and profile broadening. For Type~II and Type~III events, the state-conditioned component-flux measurements in Table~\ref{tab:mode_response} show corresponding responses in the other observing band. The $Z_0$ and PCTS variations illustrate their temporal correspondence, but these comparisons do not establish independently detected two-state switching in both bands for every event.

The confirmed Type~I--III events in the present campaign, together with the earlier bright--quiet events, show that short-time-scale switching occurred intermittently from MJD~59039 to 59399. By comparison, no minute-time-scale switching was identified in six multihour observations obtained approximately six to nine months after the outburst \citep{lbl2025}. The differences among observing campaigns are therefore more naturally attributed to the intermittent occurrence of the phenomenon, together with differences in observing epoch, duration, sensitivity, and frequency coverage, than to a permanent stabilization of the magnetosphere.

Mode changing in ordinary pulsars is generally defined as abrupt switching between two or more quasi-stable average profiles, often distinguished by changes in the intensity or phase of individual components \citep{l1971,bms1982,bmr2019}. Closely related short-time-scale profile-state changes have also been observed in radio magnetars. In PSR~J1745$-$2900, \citet{ywm2018} found frequent switching at 3.1~GHz between a double-peaked mode containing a weak leading component and a single-peaked mode, with typical mode durations of about 10~min. This behaviour occurred while the source remained in an otherwise comparatively stable long-term profile state, demonstrating that minute-scale mode switching can coexist with much slower profile evolution. In 1E~1547.0$-$5408, \citet{crh2007} observed a sudden enhancement of a low-intensity profile shoulder at 2.3~GHz that persisted temporarily and returned to its previous level after about 20~min. Although this event was not established as recurrent two-state mode changing, it provides another example of a short-lived, component-dependent profile reconfiguration in a radio magnetar. Simultaneous multifrequency observations of ordinary pulsars further show that such state changes can be broadband. For example, correlated changes in the intensity and phase of the third component of PSR~B0329+54 have been detected at 13 and 3~cm \citep{ysm2018}. Against this background, our simultaneous 2.25/8.60~GHz observations reveal several component-dependent classes involving different pulse-longitude regions, each containing events with a corresponding component response in the other band conditional on the dominant-band state classification. Type~I is dominated by the central component, Type~II involves the central and trailing components, and Type~III involves the leading and central components. The frequency dependence of the flux response also need not be identical to that of the underlying profile-state transition, as demonstrated most clearly by Type~I. Thus, its mode-changing behaviour varies both in the affected radio-beam region and in how the transition is manifested across frequency.

\subsection{Physical interpretation of mode changing and long-term evolution}

On long time-scales, the joint evolution of the profile, flux density, radio spectrum, and spin-down torque is naturally discussed in terms of a gradually untwisting post-outburst magnetosphere \citep{bel2009,pbh2012,pbh2013,kb2017}. As the current-carrying $j$-bundle contracts and the current distribution changes, the open magnetic flux, particle acceleration, and conditions for coherent radio emission can evolve in stages. This picture accounts qualitatively for the related long-term trends in the torque, radio brightness, and profile complexity, and it permits different profile components to brighten or fade as the active magnetic-flux bundles evolve.

Mode changing is commonly interpreted as switching of the pulsar magnetosphere between two or more quasi-stable configurations. Different configurations can have different current distributions, pair-production conditions, and radio-beam structures, changing the average profile, radio intensity, and, in some cases, the spin-down torque \citep{klo2006,lhk2010}. The three component-dependent classes identified in the present TMRT campaign may represent state transitions involving different regions of the radio beam. Type~I predominantly affects the central component, Type~II spans the central and trailing components, and Type~III spans the leading and central components. The corresponding dual-frequency $Z_0$ and PCTS variations are compatible with both observing bands responding to a common magnetospheric reconfiguration, while the different pulse-phase regions involved may reflect changes in distinct magnetic-flux bundles, magnetospheric azimuths, or emission heights. In this view, the long-term profile stages and minute-time-scale mode changes are two manifestations of a dynamic magnetosphere: slow evolution between observing epochs and rapid switching within individual observations.

Type~I poses a more specific challenge. The four confirmed Type~I events in the present campaign, together with the phenomenologically similar bright--quiet events in the earlier TMRT campaign, indicate that closely related short-time-scale state changes occurred at multiple epochs. In the present data, the central-component state changes are clearest at 8.60~GHz, while the corresponding 2.25~GHz component shows a state-conditioned response that varies in strength and can even have the opposite sign (Table~\ref{tab:mode_response}). A common magnetospheric state change can account for central-component variations at both frequencies, but does not by itself explain the different flux responses. The contrast between the bands is consistent with an additional frequency-dependent contribution to the observed mode response.

Coherent curvature radiation from charge bunches in a post-outburst, current-carrying $j$-bundle provides one possible spectral framework for this behaviour \citep{wzc2019}. Such emission can have a multisegment broken-power-law spectrum whose characteristic frequencies depend on particle energies, field-line curvature, and bunch geometry. A Type~I transition could shift these characteristic frequencies, producing a stronger central-component flux response at 8.60~GHz while leaving the 2.25~GHz flux relatively unchanged. This remains one possible interpretation of the frequency-dependent response.

\begin{figure}
    \centering
    \includegraphics[width=\columnwidth]{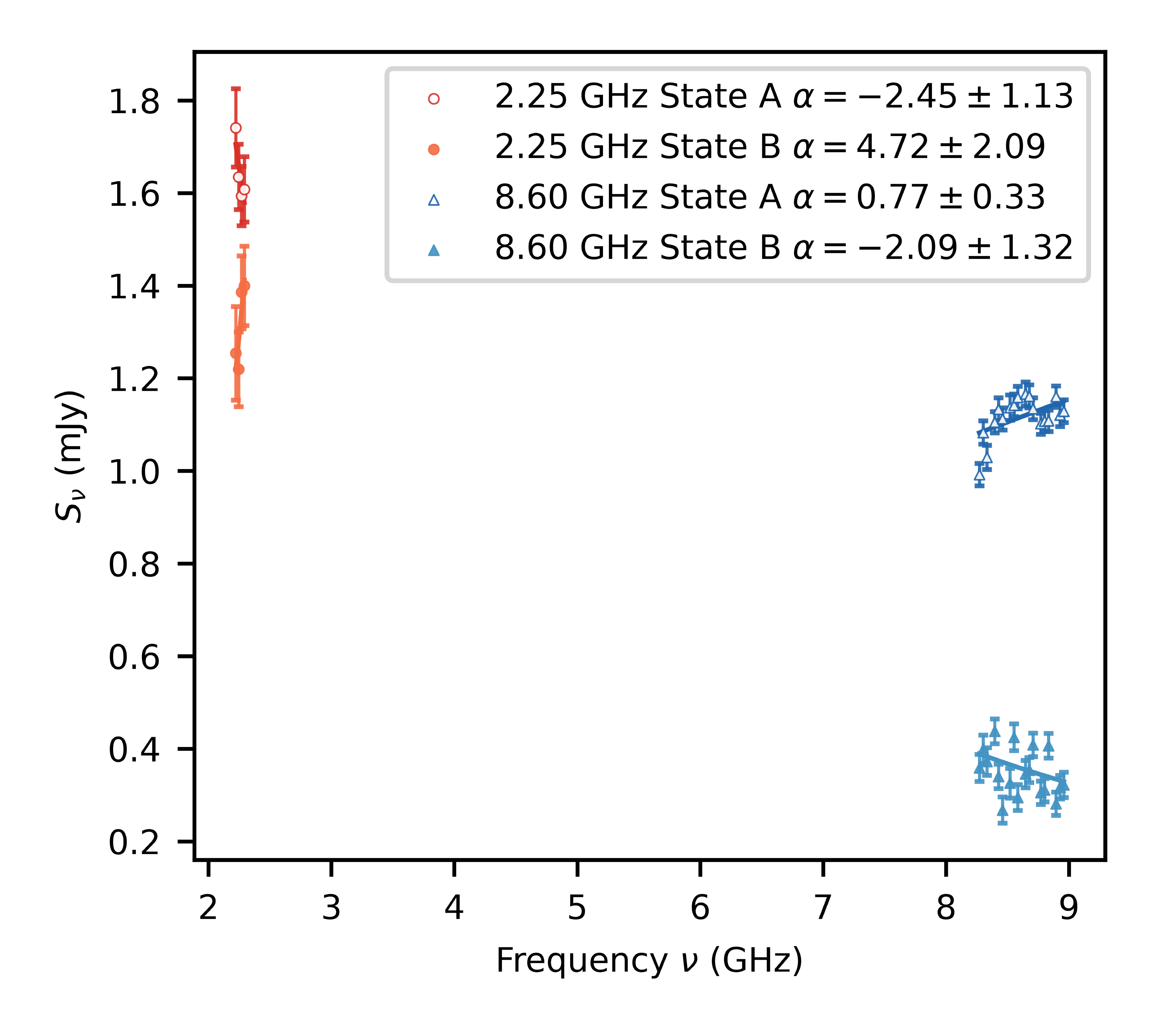}
    \caption{Mode-resolved in-band spectra of the central component on MJD~59105. Open and filled symbols show State~A and State~B, respectively, with 2.25~GHz measurements in red and 8.60~GHz measurements in blue. Error bars give the measurement uncertainties. Solid lines show the weighted power-law fits within each receiver band; the fitted indices and formal uncertainties are listed in the legend.}
    \label{fig:mode_spectra}
    \par\smallskip
    \begin{minipage}{\linewidth}\small\raggedright
    \textbf{Alt text:} Flux density versus frequency for the central component on MJD~59105, with red points near 2.25~GHz and blue points near 8.60~GHz. Open State~A points are brighter than filled State~B points in both bands. At 8.60~GHz, the State~A spectrum rises slightly with frequency, while State~B declines; the narrow 2.25~GHz band shows the opposite fitted slopes. Error bars and fitted lines accompany each group of points.
    \end{minipage}
\end{figure}

To examine whether the frequency-dependent Type~I flux response is accompanied by an in-band spectral change, we measured the mode-resolved spectra of MJD~59105, the clearest event of this type (Fig.~\ref{fig:mode_spectra}). Each state provides four fitted points with central frequencies of 2.222--2.292~GHz and 18 fitted points from 8.271 to 8.958~GHz.

At 8.60~GHz, State~A is brighter and has a flatter fitted in-band trend, with $\alpha=+0.77\pm0.33$ compared with $-2.09\pm1.32$ in State~B. At 2.25~GHz, the formal fits give $\alpha=-2.45\pm1.13$ and $+4.72\pm2.09$ for States~A and B, respectively. The usable 2.25~GHz bandwidth is only about 100~MHz, and its four closely spaced points make these slopes sensitive to binning and individual measurements. We therefore show the 2.25~GHz spectra but do not use their fitted indices for physical interpretation.

The flatter 8.60~GHz trend in State~A is consistent with a spectral contribution to the frequency-dependent flux response of Type~I, but does not uniquely identify its physical origin. In addition to a shift of the characteristic frequencies within the coherent-curvature-radiation framework, the observed spectral difference could arise from changes in the relative contributions of profile components with different spectral properties, as suggested by the component-dependent spectral behaviour reported by \citet{lbl2025}. Given the in-band spectral-index uncertainties and possible correlations among frequency channels, this comparison is not a statistically decisive test.

The present observations do not permit a quantitative fit of the coherent-curvature-radiation model. Two separated observing bands cannot locate the peaks or breaks of the true broadband spectrum, nor can they independently constrain the bunch scale, opening angles, field-line curvature, or particle-energy distribution. We therefore do not infer physical parameters from the broken-power-law framework; we use it only as a possible explanation for the frequency-dependent flux response of Type~I \citep{wzc2019}. The time resolution of the segmented timing analysis is also insufficient to determine whether individual minute-time-scale transitions are accompanied by instantaneous torque changes. Strictly simultaneous broadband observations, together with mode-resolved polarimetry, single-pulse analysis, and higher-time-resolution torque measurements, will be required to connect local changes in the radio-emitting region to the global magnetospheric state.

\section{Conclusions}
\label{sec:conclusions}

We have presented 65 new epochs of simultaneous 2.25 and 8.60~GHz observations of Swift~J1818.0$-$1607 obtained with the TMRT from MJD~59104 to 59427, extending our earlier monitoring of its 2020 radio-active phase. Together, the two campaigns reveal pronounced stage-dependent evolution in the pulse profile, spin frequency, flux density, and two-point spectral index. The outer profile components evolve strongly, with the late-time increase in profile width being driven mainly by enhanced leading emission. The extended timing baseline confirms the continued decrease of the spin frequency; we summarize its changing slope with four descriptive intervals. The combined data give an average spin-down rate of $\dot{\nu}=-1.65(4)\times10^{-11}\,\mathrm{s^{-2}}$, corresponding to an effective characteristic spin-down time-scale of $0.71(2)$~kyr over the analysed interval. The radio flux density first rises and then declines overall, while retaining substantial nonmonotonic fluctuations and temporary recoveries. The initially strongly negative two-point spectral index subsequently shows repeated decreases and recoveries, with values greater than $-1$ at 41 of the 54 later epochs with usable detections in both bands. The broad association among profile complexity, radio brightness, and spin-down behaviour suggests that these observables respond to the same evolving post-outburst magnetosphere, but with different amplitudes and characteristic time-scales.

We identify 15 confirmed events in three component-dependent classes of minute-time-scale mode changing, with two additional visually suggestive events retained as tentative. Type~I is dominated by changes in the central component and resembles the bright--quiet switching detected in our earlier TMRT campaign, demonstrating that this behaviour recurred during the post-outburst radio-active phase. The central-component state changes are most conspicuous at 8.60~GHz, while the 2.25~GHz response varies among events. Type~II involves the central and trailing components, with enhanced trailing emission providing the clearest distinction between the two states. Type~III involves the leading and central components and is characterized by enhanced leading-side emission and profile broadening. All three classes include events with a corresponding component response in the other observing band, measured conditional on the dominant-band state classification. These responses do not establish independent two-state detections in both bands for every event, and their strengths and directions vary among events. The short-time-scale state switching of Swift~J1818.0$-$1607 is therefore component dependent and cannot be described by a single uniform profile transition.

The combined long- and short-time-scale behaviour is consistent with a post-outburst magnetosphere undergoing both gradual reconfiguration and rapid switching between quasi-stable states. The different pulse-phase regions involved in Type~I--III may reflect changes in distinct magnetic-flux bundles, magnetospheric azimuths, or emission heights. For the representative Type~I event on MJD~59105, the brighter 8.60~GHz state also shows a flatter in-band spectral trend, consistent with a spectral contribution to its strongly frequency-dependent flux response. Although a coherent-curvature-radiation framework can accommodate such behaviour through changes in the characteristic frequencies of the radio spectrum, the present two-band observations do not permit a quantitative test of this interpretation. Strictly simultaneous observations with broader frequency coverage, together with mode-resolved polarimetry, single-pulse analysis, and higher-time-resolution torque measurements, will be required to determine how profile-state changes, spectral evolution, and the global magnetospheric state are physically connected.

\section*{Acknowledgements}
This work was supported in part by the National SKA Program of China (Grant No.~2020SKA0120104), the National Key Research and Development (R\&D) Program of China (Grant No.~2022YFA1603104), the National Natural Science Foundation of China (Grant Nos.~U2031119, 12041301, and 1257030658), and the Scientific Research Foundation of Hubei University of Education for Talent Introduction (Grant No.~ESRC20240007). We thank Dr~Jie Liu for constructive suggestions on revising this manuscript. We thank the staff of the TMRT for their support with the observations. Generative AI tools were used to assist with language editing and code-based figure preparation. All scientific content, analyses, numerical results, figures, interpretations, and final text were checked and approved by the authors, who take full responsibility for the manuscript.

\section*{Data Availability}
The data underlying this article will be shared on reasonable request to the corresponding authors.


\bsp
\label{lastpage}
\end{document}